# MODIFIED TEMPORAL KEY INTEGRITY PROTOCOL FOR EFFICIENT WIRELESS NETWORK SECURITY

M. Razvi Doomun*, KM Sunjiv Soyjaudah

*Faculty of Engineering, University of Mauritius, Reduit*
r.doomun@uom.ac.mu, ssoyjaudah@uom.ac.mu

Abstract: Temporal Key Integrity Protocol (TKIP) is the IEEE TaskGroupi's solution for the security loop holes present in the already widely deployed 802.11 hardware. It is a set of algorithms that wrap WEP to give the best possible solution given design constraints such as paucity of the CPU cycles, hardwiring of the WEP encryption algorithm and software upgrade dependent. Thus, TKIP is significantly more difficult and challenging to implement and optimise than WEP. The objective of this research is to examine the cost/benefit of TKIP security mechanisms and optimise its implementation to reduce security overhead for better performance. We propose a modified TKIP (MoTKIP) with improved packet encapsulation and decapsulation procedure that reduces computation and packet overhead in classic TKIP substantially and optimises total wireless network throughput rates.

## 1 INTRODUCTION

Most current wireless networks are based on widely adopted IEEE 802.11 standard (Stallings, 2003). Unfortunately the initial security specification, called Wired Equivalent Privacy (WEP), in this standard has been proved insecure and is thus inadequate for protecting a wireless network from eavesdropping and other attacks (Borisov, Goldberg, & Wagner, 2001). TKIP mechanism is a transition method used to strengthen security of IEEE 802.11 WLAN and is implemented through software upgrades using RC4 of WEP as its core, but introduces changes in the areas of message integrity, IV creation, and key management. However, the additional computation complexity and overhead that TKIP key mixing and encapsulation process impose have received little research attention. Even so, cryptographic review thus far suggests it achieves its fundamental design goals (Al Naamany, Shidhani & Bourdoucen, 2006).

The TKIP encapsulation process and message integrity check increase the size of the transmitted packets, which in turn lower the effective bandwidth and increase the communication cost also. Consequently, reducing security overhead and power consumption are current research challenges to wireless security designers so as to extend the operating lifetime of battery powered mobile devices. The aim of this work is to study TKIP and optimally trade between security overhead and power consumption to design a modified TKIP (MoTKIP) with low packet encapsulation and decapsulation overhead. The rest of the paper is organized as follows: Section 2 gives a succinct account of literature and related work in the field of wireless security optimization. We present a brief overview of the TKIP algorithm in Section 3. The MoTKIP packets and optimised TKIP key mixing are described in Section 4. Then, in section 5 a security overhead comparison is performed. This is followed by concluding remarks in Section 6.

## 2 RELATED WORK

In the literature (Prasithsangaree & Krishnamurthy, 2004), most efficient wireless security algorithms are designed based on models which do not take into account the security performance together with the computational overhead. Research by Jones *et al.* (Jones, Sivalingam, Agrawal, & Chen. 2001) and Lettieri *et al.* (Lettieri & Srivastava, 1999) have shown that one of the main causes of unnecessary energy consumption is overhead of security and communication protocol over a wireless channel. Two basic principles that have been suggested to achieving an energy efficient system are to avoid unnecessary actions and reduce the amount of data traffic (Havinga & Smit, 2001). However, most existing wireless security protocols use resources avidly and limit the efficient use of wireless node resources in several ways. Wireless security protocols significantly increase the amount of overhead required to secure the network, thereby decreasing the data rates of wireless links because additional traffic is usually needed for authentication or verification services (Thomas, Al-Begain & Hughes, 2005). Another prior work from Ganesan *et al.* (Ganesan, Venugopalan, Peddabachagari, Dean, Mueller & Sichitiu, 2003) assesses the feasibility of different encryption schemes for a range of embedded architectures using execution time overhead measurements. Potlapally *et al.* (Potlapally, Ravi, Raghunathan & Jha, 2003.) investigated energy consumption of different ciphers on the Secure Sockets Layer. Consequently, our work consolidates all earlier work on 802.11 wireless securities and adds to those research works to fill this void of, investigating and answering the question of designing low overhead TKIP encryption for existing 802.11 hardware

## 3. OVERVIEW OF TKIP

TKIP increases the size of the initialisation vector (IV) used in the encapsulation process to an effective 48 bits. This significantly decreases the probability of an IV reuse by increasing the size of possible IVs to $2^{48}$ as opposed to $2^{24}$ possible WEP IV values. Increasing the IV length also addresses WEP's weak key vulnerability. It achieves this by implementing a very innovative way of splitting the IV into two parts. The first 16 bits of the least significant part of the IV are padded to create a 24-bit IV in a way that avoids the use of weak keys. This process is called per-packet key mixing. This IV is joined to a mixed key that is calculated using the remaining most significant 32 bits of the TKIP IV as well as the MAC-address of the wireless LAN card to generate the key. It ensures that every packet has a different set of IVs. Thus, one of the main problems of the WEP algorithm is solved, namely that every station belonging to the network is using the same key for encrypting data.

TKIP uses a temporal key (TK) and the packet sequence number to arrive at a per-packet key and IV. The temporal key is a 128-bit shared secret between transmitter and receiver that has a fixed lifetime. TKIP uses a two-phase key mixing operation to derive the unique per-packet key stream, and each phase fixes one particular flaw in WEP. Phase 1 eliminates the same key from use by all links and Phase 2 de-correlates the public IV from known per-packet key.

Phase 1 mixes 128-bit Temporal Key (TK) with the first 4 bytes of IV and the sender's 48-bit MAC address, and generates an intermediate key P1K. In this step, the 48-bit MAC address of the transmitter is iteratively XORed with the 128-bit TK. Each byte of the result is used to index an S-box, an invertible non- linear substitution table, to produce the 80-bit intermediate key P1K. This intermediate key is computed once every $2^{16}$ packets and is likely to be cached. For performance optimization, intermediate key P1K is computed only when the temporal key is changed/updated and most of the time its value is saved on memory. By mixing the MAC address into the temporal key, Phase 1 ensures that various stations using the same temporal key generate different key streams. Hence, this prevents key stream reuse due to cross-station IV collision.

Phase 2 takes input P1K [80 bits] with TK (128 bits) and the last 2 bytes of IV to generate a unique 128-bit RC4 key, also known as WEP seed. It employs S-box substitution, rotate operation and addition operation to generate the 128-bit per-packet RC4 key. This decouples the known association between IV and the key, thus preventing exploiting weak keys to recover TK. However, the RC4 key has an internal structure that must conform to the WEP specification for compatibility. That is, the first 24-bits are used to convey the WEP IV and the last 104 bits convey the WEP base key, as the existing WEP hardware expects to concatenate a base key and IV. This is accomplished by assigning the 8 most significant bits of the packet sequence number to the first and second bytes of the WEP IV, and the least significant sequence number bits to the third IV byte. The first 3 bytes of Phase 2 output correspond exactly to the WEP IV and the last 13 bytes to the WEP base key. Precisely, the first and third byte of RC4 per packet key comes from the lower 16-bit of the IV. The second byte is a repeat of the first byte, except that bit 5 is forced to 1 and bit 4 is forced to 0. Phase 2 applies this masking off specific bits of the second IV byte to prevent the WEP per-packet key concatenation from producing the known RC4 weak keys (Paul & Preneel, 2004.). Finally, the RC4 key is used to generate the key stream, which is then XORed with the plaintext for encryption.

TKIP security mechanism takes active countermeasures when two MIC failures are detected in less than one minute. The countermeasures generally consist of re-keying the connection and notifying the network administrator. When the receiver encounters two packets with invalid MICs within one minute, it believes to be under an attack, and disassociates its clients and waits for a minute before continuing operation. MIC is checked last in the TKIP decapsulation process. Any frame that does not have a valid ICV and TSC are discarded before the MIC is verified. The ICV ensures that noise and transmission errors do not erroneously trigger the countermeasures. In order to minimize the risk of false alarms, the rule is that the MIC shall be verified after the CRC, IV and other checks have been performed (IEEE Standard for Information Technology, 2004). Another countermeasure is that, if a new temporal key cannot be established before the full 16-bit space TSC is exhausted, then TKIP protected communications will cease. For key refresh failure, the implementation halts further data traffic until rekeying succeeds, or disassociates.

In short, TKIP is designed in such a way that security completely relies on the secrecy of all the packet keys (Moen, Raddum & Hole, 2004). Even if one packet key is lost to the attacker, it is easily possible to find the MIC key. Similarly, if two packets with same IV are disclosed, an attacker can do anything for the duration of the current temporal key. To avoid replay attacks, the sequence counter is simply implemented such that no IV value, which once has been received, will be allowed. The only problem of this approach is introduced by the fact that IEEE 802.11 allows burst-acknowledgements, which indicates that up to 16 packets could be sent at once and then be acknowledged by just a single packet. Consequently the sequence counter has to remember the last 16 IV values to guarantee that all packets have been correctly received.

## 4 MODIFIED TKIP

As shown in Figure 1, the second phase of TKIP key mixing function reuses the 80-bit TKIP-mixed Transmit Address and Key (TTAK) or phase 1 key (P1K) with MAC Protocol Data Units (MPDUs) associated with the same 32-bits upper IV part, Temporal Key (TK) and Transmitter Address (TA) for the next consecutive $2^{16}$ packets. Hence, the 32-bit high IV becomes known to the receiver when the first encrypted packet is transmitted and this part is cached since it is static for the subsequent $2^{16}$ packets.

The second phase mixes the output of the first phase with the TK and monotonically increments 16-bit low IV part counter (i.e. 0x0000-0xFFFF) to produce the final WEP seed, also called the per-packet key. Since the knowledge of 32-bit high IV and the future sequence of 16-bit low IV is also known to the receiver, it is not necessary to send the full 48-bit extended IV as redundancy again in each packet. Thus, the cached 80-bits TTAK derived from the IV in the first packet at the transmitter will also be that same input to the second phase mixing of the receiver and automatically the next 16-bit IV counter it is just a unit increment of the previous one. Since the IV counter is predictable, phase 2 can be computed in advance while waiting for the next packet(s) to arrive at the receiver. Therefore, in the new Modified TKIP (MoTKIP) frame format, the redundant 4-bytes extended IV is removed from the packet load for packets ranging from the $2^{nd}$ to the $(2^{16})^{th}$ packet. We use the standard code algorithm in (IEEE Standard for Information Technology, 2004) to optimised TKIP key mixing phase.

The function MK16 constructs a 16-bit value from two 8-bit inputs as MK16(X,Y) = (256*X) + Y. The phase 1 output stays the same for $2^{16}$ (i.e. 65, 536) consecutive frames from the same TK and TA. Cheap CPU operations common on 802.11 devices, such as the exclusive-OR operation ($\oplus$), the addition operation (+), the AND operation (&), the OR operation ( | ), the right bit shift operation (>>), rotate and table look-ups are used in phase 1 and phase 2 key mixing.

Figure 1 also illustrates the general procedure for MoTKIP. For MoTKIP encapsulation process, another major change in TKIP frame format is implemented by calculating the MIC over IV also; and only for first packet transmission the Extended IV XORed with session key is concatenated and sent in the packet. In addition, the MoTKIP encapsulation uses special flag bits for specific control purpose. At the start of secure communication, the transmitter or sender computes a keyed cryptographic message integrity code, or the MIC, over the MSDU source and destination addresses, the priority bits, the MSDU plaintext data and the 48-bit Extended IV also. MoTKIP appends the computed MIC to the MSDU data prior to fragmentation into MPDUs. The receiver obviously has to verify the MIC after decryption with Extended IV, ICV checking, and reassembly of the MPDUs into an MSDU. Invalid MIC naturally leads to discarding of corresponding MSDUs, and this defends against forgery attacks and replay attacks.

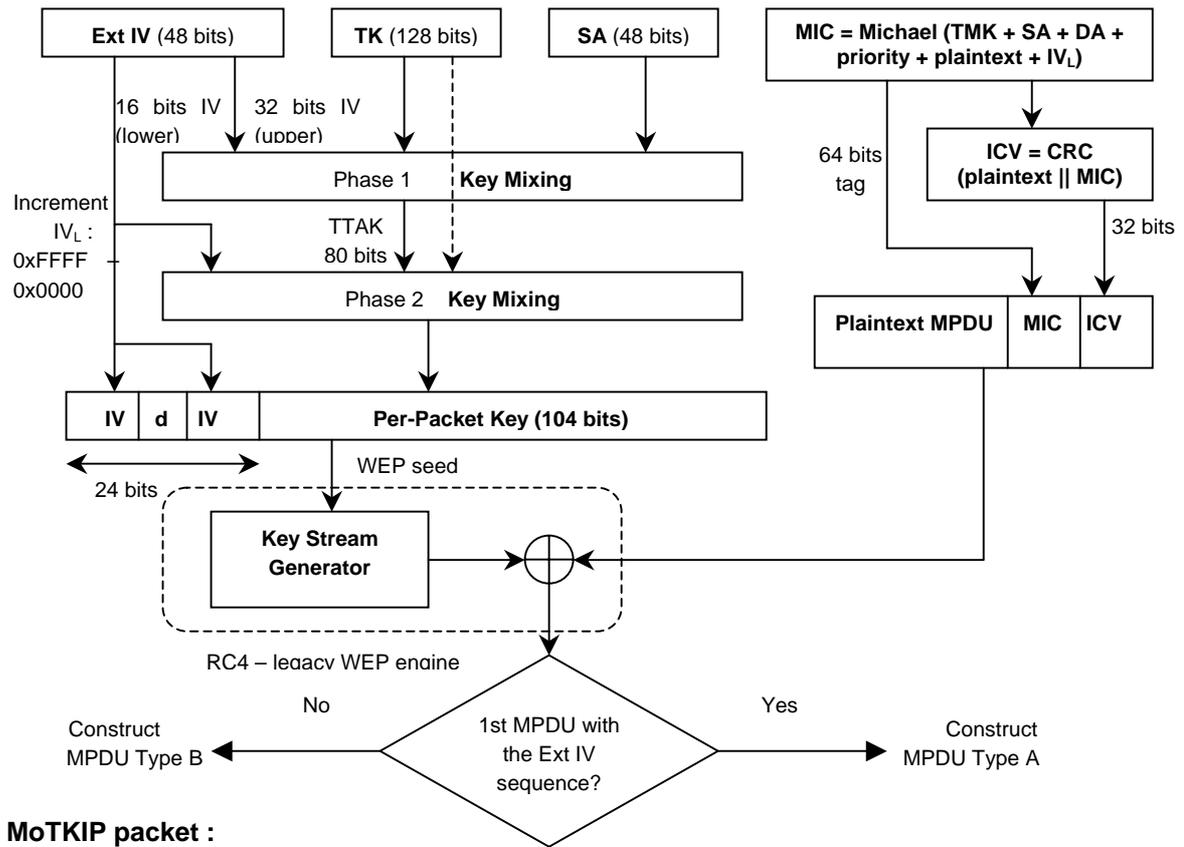

**First MoTKIP packet :**

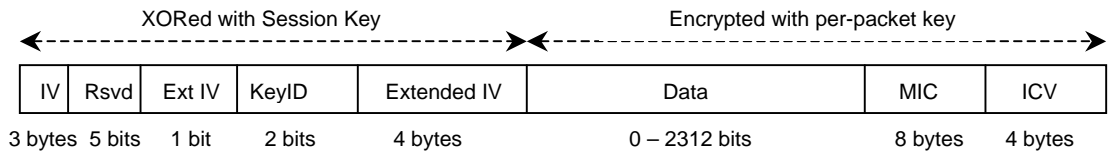

**2nd and Subsequent MoTKIP packet :**

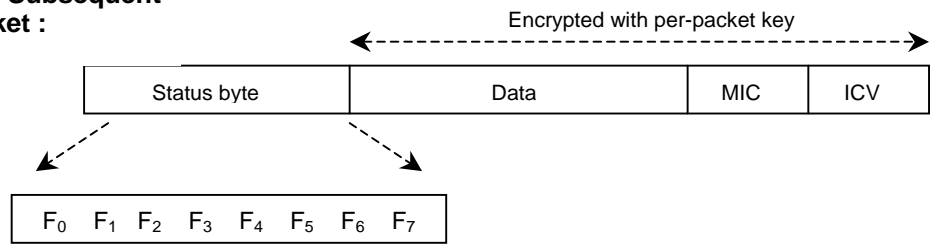

$F_0$ : Indicate Ext IV absent/present
$F_1$ : Indicate same $IV_H$ for phase 1 key mixing
$F_2$ : Indicate $IV_L$ for current packet decapsulation is unit increment of previous $IV_L$
$F_3$ : Indicate if MSDU fragmented
$F_4$ : Indicate if MPDU belongs to same MSDU
$F_5$ $F_6$ $F_7$ : Rsvd

Figure 1. Modified TKIP (MoTKIP) procedure with low overhead MPDU.

MoTKIP uses the cryptographic key mixing function to combine a temporal key (TK), transmitter address (TA), and the extended IV into the WEP seed similar to the classic TKIP. At the start of the session, for the first data packet, the 48-bit extended IV XORed with a session key is appended to the encrypted data.

However, the new MIC for each MoTKIP packet is calculated using the IV, source and destination addresses and data payload at the transmitter, and the same is recalculated at the receiver to detect replay attacks.

In the traditional approach, the key reason why the IV is transmitted in the clear is because the 802.11 standard assumes that an adversary does not gain any useful information from its knowledge. The IV is meant to introduce randomness to the key, and appending the clear IV in the transmitted packet helps the receiver to decrypt the information sent from the transmitter station. However, it has been proved that various types of attacks are possible using the IV knowledge as described in the literature (Walker, 2004; Borisov, Goldberg & Wagner, 2001).

In our new approach to strengthen the TKIP security, for the first packet transmitted when a session starts between STA and AP, the extended IV is encrypted with a session key. Hence, first MoTKIP packet sent with encrypted IV uses **C = [ Ks $\oplus$ IV, P $\oplus$ RC4 ( IV, TK, SA) ]**. The MoTKIP MIC is computed over: the MSDU destination address (DA); the MSDU source address (SA); the MSDU priority (Reserved for future use); the entire unencrypted MSDU data (payload) and the unencrypted entire IV also. The DA, SA, clear extended IV, 3 octets reserved to 0 and a one octet priority field are used for calculating the MIC and are not transmitted. The TKIP decapsulation mechanism is composed of several sub processes all working continuously to provide the decrypted packets. The process of decapsulating the first MoTKIP packet is the opposite of the encapsulation process of the first packet with the addition of the integrity checks. The MoTKIP decapsulation is designed to be the least computationally intensive. Since the predictable rule for sequencing the Extended IV for subsequent TKIP packets (from $2^{nd}$ to $65536^{th}$ packet) is to increment the low 16-bit IV part monotonically from 0xFFFF to 0x0000, the 48-bits Ext IV is not included in these MoTKIP packets. Instead, a status byte is appended as shown in Figure 1.

## 5 PERFORMANCE

The experimental test setup consisted of establishing and testing secure communication performance via 802.11b wireless network cards and an access point. The nominal data rates were 11Mbps. Encrypted files are transmitted from clients to server. As the server is receiving data packets, it checks for the status byte header to identify the proper MoTKIP algorithm for decryption. Several statistical experiments were performed to verify the MoTKIP operation and WLAN performance throughput.

Table 1 shows the relative security overhead imposed by WEP, classic TKIP and MoTKIP when applied to all traffic types and also includes the open security reference benchmark. Wireless device battery may experience reduced efficiency when several intensive operations (power hungry), such as key mixing, MIC calculation and RC4 processing, occur simultaneously. As expected, 128-bit WEP encryption imposes the least overhead whilst TKIP demands significant use of the available wireless bandwidth because of extended IV and MIC, placing substantial burden on WLAN performance. While classic TKIP is predictably more intensive than WEP, the processing overhead of MoTKIP is substantially less than classic TKIP. MoTKIP outperforms TKIP in terms of percentage computational overhead and energy consumption efficient by 25–35%. The throughput of MoTKIP measured in the simulation tests clearly shows MoTKIP to be less bandwidth intensive. In terms of overhead percentages associated with each encryption schemes compared to no encryption: 128-bit WEP incurs 1.0 – 1.4% throughput degradation; 128-bit classic TKIP brings 2.4 – 2.6% overhead increase; while MoTKIP incurs only 1.9- 2.1% overhead that lowers actual data throughput rate.

Table 1: 802.11 WLAN throughput v/s encryption scheme

| Encryption schemes | Average throughput (Kbps) |
|---|---|
| No encryption | 8695 |
| WEP-128 bits | 8580 |
| TKIP-128 bits | 8254 |
| MoTKIP – 128 bits | 8472 |

Table 2: Processing Time by encryption schemes without transmission

| Encryption schemes | Processing Time (ms) |
|---|---|
| No encryption | 194 |
| WEP-128 bits | 298 |
| TKIP-128 bits | 625 |
| MoTKIP – 128 bits | 501 |

Table 2 shows the time consumed for iteration over a 5MB file. The time consumed is proportional to the complexity of encryption operations and the amount of data transmission. From these results it can be concluded that MoTKIP should be used wherever implementable due to its lower overhead and enhanced security in comparison with classic TKIP.

# 6 CONCLUSION

In this paper, we study the performance overhead caused by TKIP compared to our proposed MoTKIP. The latter is a better optimised intermediate solution than TKIP and it also discards all known attacks from WEP architecture while preserving the RC4 algorithm to ensure compatibility. Lightweight MoTKIP security scheme is used to minimize the impact of overhead on network and energy resources. Simulation results demonstrate the effectiveness of our approach with higher network throughput gain for embedding minimal information into the MoTKIP encapsulated packets. It considerably decreases key mixing complexity of the encryption while making it securely more robust and efficient.